\documentstyle[12pt]{article}

\def\be{\begin{equation}}
\def\ee{\end{equation}}

\begin{document}
\title{Gauging of 1d-space translations for nonrelativistic point 
particles}
\author{P.C. Stichel\thanks{e-mail: hanne@physik.uni-bielefeld.de}\\An der Krebskuhle 21\\
D-33619 Bielefeld, Germany}
\date{}
\maketitle

\begin{abstract}
Gauging of space translations for nonrelativistic point particles in one
dimension leads to general coordinate transformations with fixed
Newtonian time. The minimal gauge invariant extension of the particle
velocity requires the introduction of two gauge fields whose minimal
self interaction leads to a Maxwellian term in the Lagrangean. No
dilaton field is introduced. We fix the gauge such that the residual
symmetry group is the Galilei group. In case of a line the two-particle
reduced Lagrangean describes the motion in a Newtonian gravitational
potential with strength proportional to the energy. For particles on a
circle with certain initial conditions we only have a collective 
rotation with constant angular velocity.
\end{abstract}

\medskip
~~~PACS: 03.20.+i; 11.15.-q

\section{Introduction}

Gauging a symmetry means to introduce space-time dependent functions 
instead of the constant parameters of a global symmetry group. The
original action, invariant with respect to the global symmetry group,
will in general not be gauge invariant. In order to cancel the terms
which violate the invariance one has to introduce compensating fields
as functions of space but also of time. These gauge fields have to obey
suitable transformation laws with respect to the now local symmetry.

The most important interactions in current physics are gauge interactions.
In elementary particle physics internal symmetry groups are gauged
(Quantumelectrodynamics, Quantumchromodynamics, Standard Model). In 
general relativity, the space-time symmetry group of special relativity
(Poincar\'e group) is gauged. Gauge interactions are also of importance 
in condensed matter physics: the fractional quantum hall effect may be 
described by an abelian gauge field with a Chern-Simons term minimally
coupled to charged matter. Therefore the ``gauge principle'' may be
called the leading principle for the construction of fundamental interactions
in physics \cite{Mills}.

What about gauging the Galilei group in nonrelativistic physics? In a
recent paper by De Pietri et al.~\cite{Pietri} this task has been taken
up for point particles in (3+1)-dimensions. The authors of \cite{Pietri}
started with the nonrelativistic limit of general relativity and threw
away all fields not coupled to matter in this limit. Our work will be
distinct from \cite{Pietri} in two respects:
\begin{itemize}
\item[i)] We ask for the smallest number of gauge fields doing the job
with a minimal coupling Lagrangean,
\item[ii)] we don't start with the nonrelativistic limit of a relativistic
theory.
\end{itemize}
In this letter we treat the simplest example: point particles moving in
one dimension. The underlying global transformations are space translations
whose gauging leads to general coordinate transformations at fixed
Newtonian time. For particles on a line we will show in section 2 that
the minimal coupling for this example consists of only two gauge fields
with an Maxwellian interaction term. In particular we don't introduce an
additional dilaton field as in (1+1)-gravity \cite{Cangemi}. The arbitrary
gauge function in the solution for the gauge fields will be chosen such,
that the residual symmetry group of the action will be the Galilei group
and the boundary term for field variations in the action vanishes (section 3).
In section 4 we derive the reduced Lagrangean for one and two particles
respectively and discuss the solutions of their equations of motion. In
section 5 we extend our results to particles on a circle. With certain
initial conditions relative motions of particles are absent. We obtain
a collective rotation with constant angular velocity only.

\section{Minimal-coupling Lagrangean}

Let us start with N nonrelativistic particles in free motion on a line
$({\bf R}^1)$ described by the Lagrangean\footnote{For reasons of simplicity
we give all particles the same mass m=1 in appropriate units.}
\be
L_0 = {1 \over 2} \sum^N_{\alpha = 1} (\dot{x}_\alpha (t))^2
\ee
The equations of motion (EOM) following from (1)
\be
\ddot{x}_\alpha = 0
\ee
are invariant with respect to global Galilei-transformations
\be
(x,t) \to (x^\prime, t^\prime)
\ee
with
\be
x^\prime = x + a + vt
\ee
and
\be
t^\prime = t + b
\ee
where the parameters $a$, $v$ and $b$ take values in ${\bf R}^1$. Now we
generalize (4) to a local transformation, given in infinitesimal form by
\be
\delta x = a (x,t)
\ee
where $a (x,t)$ is an arbitrary, twice differentiable and bounded function
of its arguments. Eq. (6) describes local translations (including local
boosts). Time translations are not considered for the moment.\\
Obviously the EOM (2) are not invariant with respect to the transformation 
(6). In order to repair that, we introduce two gauge fields $h (x,t)$ and
$e (x,t)$ and replace $\dot{x}$ in (1) by the function\footnote{Our
procedure differs from the corresponding one in [2] applied to one space
dimension. In [2] $L_0$ would be replaced by a polynomial of second order
in $\dot{x}$ requiring three gauge fields instead of two.} $\xi$
\be
\xi = h (x,t) \dot{x} + e (x,t).
\ee
Invariance of $\xi$ with respect to (6) requires the following transformation
rules for the gauge fields
\be
\delta h = - h \partial_x a~,~\delta e = - h \partial_t a
\ee
where we defined
\be
\delta f (x,t): = f^\prime (x + \delta x, t) -f (x,t).
\ee
Therefore, (7) supplies the minimal gauge invariant extension of
$\dot{x}$. Now $L_0$ in (1) has to be replaced by
\be
L_{\rm matter} = {1 \over 2} \sum^N_{\alpha = 1} (\xi_\alpha (t))^2
\ee
with
\be
\xi_\alpha (t): = h (x_\alpha (t), t) \dot{x}_\alpha (t) + e(x_\alpha
(t), t)
\ee
We must supplement (10) by an invariant (or quasi-invariant) Lagrangean
${\rm L_{\rm field}}$ describing the self-interaction of the gauge fields.
Let us define a field strength $F$
\be
F: = {1 \over h} (\partial_t h - \partial_x e)
\ee
From the definition (9) we obtain easily the commutator between $\partial$ 
and any partial differentiation $\partial \in (\partial_t, \partial_x)$
\be
\delta \partial f = \partial \delta f - (\partial a) \partial_x f
\ee
We infer from (13) and (8) that our field strength $F$ is gauge invariant
\be
\delta F = 0
\ee
Therefore, any integral of the form
\be
\int_{{\bf R}^1} d \mu_t (x) K (F (x,t))
\ee
with the invariant measure
\be
d_{\mu_t} (x): = h (x,t) dx
\ee
is a candidate for ${\rm L_{\rm field}}$.
The simplest, nontrivial example for the arbitrary function $K$ is given by
a quadratic
\begin{eqnarray*}
K(Z) = Z^2
\end{eqnarray*}
With this Maxwellian choice for ${\rm L_{\rm field}}$ our action takes the
form
\be
S = \int dt ({\rm L_{\rm matter}} + {\rm L_{\rm field}})
\ee
with
\be
{\rm L_{\rm field}} = {1 \over {2g}} \int dx h (x,t) F^2 (x,t)
\ee
where $g$ is a coupling strength.\\
By varying $S$ with respect to $x, h$ and $e$ we get the EOM
\be
\dot{\xi}_\alpha + \xi_\alpha F_\alpha = 0
\ee
\be
\partial_t F + {1 \over 2} F^2 = g \sum_\alpha \xi_\alpha \dot{x}_\alpha
\delta(x - x_\alpha)
\ee
\be
\partial_x F = -g \sum_\alpha \xi_\alpha \delta (x - x_\alpha)
\ee
In order that the boundary term arising in the derivation of (21) vanishes
and the integral (18) for ${\rm L_{\rm field}}$ exists, our gauge fields
have to satisfy the following boundary conditions (A) at spatial infinity:
\begin{itemize}
\item[(A1)] $e (x,t)$ and $h (x,t)$ are finite at $x = \pm \infty$,
\item[(A2)] $F (x,t)$ vanishes at $x = \pm \infty$.
\end{itemize}

\section{Solutions of the field equations and gauge fixing}
The solution of (21) satisfying (A2) is given by $(\epsilon (x):= x/\vert x
\vert)$
\be
F(x,t) = - {g \over 2} \sum_\alpha \xi_\alpha (t) \epsilon (x - x_\alpha (t))
\ee
with the constraint
\be
\sum_\alpha \xi_\alpha (t) = 0
\ee
With (19), (22) and (23) the EOM (20) is satisfied identical.\\
From the expression (22) for the field strength $F$ and its relation to
the gauge fields (12) we obtain $h$ and $e$ only modulo an arbitrary gauge
function $\Lambda (x,t)$
\be
h (x,t) = \partial_x \Lambda (x,t)
\ee
\begin{eqnarray}
e (x,t) & = & \partial_t \Lambda (x,t) + \nonumber\\
& + & {g \over 2} \sum_\alpha \xi_\alpha (t) \epsilon (x - x_\alpha (t))
(\Lambda (x,t) - \Lambda_\alpha (t))
\end{eqnarray}
Gauge fixing means to choose an appropriate function (class of functions)
for $\Lambda$. Thereby we have to keep in mind, that the relation between
$\xi_\beta$ and the particle variables $\{ x_\alpha, \dot{x}_\alpha \}$
is gauge dependent\footnote{This situation reminds us of general relativity.}.
A physical choice for $\Lambda$ should fulfill two conditions:
\begin{itemize}
\item[i)] The $x_\alpha (t)$ describe particle motion in an inertial frame.
Therefore, the symmetry remaining after gauge fixing (residual symmetry) is
the Galilean symmetry in (1+1)-dimension.
\item[ii)] The boundary condition (A1) is satisfied.
\end{itemize}
We try to satisfy the first condition with the ansatz
\be
\Lambda (x,t) = x - (a + vt)
\ee
because due to (8) and (13) we have $\delta \Lambda = 0$
and therefore with (26)
\be
\delta x = \delta a + t \delta v
\ee
i.e. the residual symmetry is given by the Galilei group\footnote{Time 
translations will be considered in section 4.}.
In section 4 we will show, that the resulting particle EOM describe
particle motion in an inertial frame at least for $N=1$ and $N=2$.
Now the second condition (A1) is satisfied too. From (23) - (26) we infer
\be
h (x,t) = 1
\ee
and
\be
e (x,t) \stackrel{\vert x \vert \to \infty}{\longrightarrow}
- v - {g \over 2} \epsilon (x) \sum_{\alpha} \xi_{\alpha} x_\alpha
\ee

\section{Reduced particle Lagrangean}

With the results of section 3 the following equations remain for the 
determination of the particle trajectories $x_\alpha (t)$:
\begin{description}
\item{i)} \be
\xi_\alpha = \dot{x}_\alpha - v + {g \over 2} \sum_\beta \xi_\beta 
\vert x_\alpha - x_\beta \vert
\ee
determines $\xi_\alpha$ in terms of the $\{ x_\beta\}$. It is obtained
from (7) and (24) - (26).
\item{ii)}
\be
\dot{\xi}_\alpha = {g \over 2} \sum_\beta \xi_\alpha \xi_\beta \epsilon
(x_\alpha - x_\beta)
\ee
obtained from (19) and (22),
\item{iii)}
\be
\sum_\alpha \xi_\alpha (t) = 0\ .
\ee
\end{description}
In our gauge the total canonical particle momentum $P$ is given by
\be
P = \sum_\alpha \xi_\alpha
\ee
which is conserved due to (31)\footnote{Our gauge fields carry no
dynamical degrees of freedom. Therefore, the field contribution to
the momentum vanishes.}. It even has to vanish according to the constraint
(32).
Let us now discuss (30) - (32) for different particle numbers $N$.

\medskip
\noindent
{\underbar{N = 1}

\smallskip
\noindent
We have $\xi = \dot{x} - v$ and $\xi = 0$ i.e.
\be
\ddot{x} = 0
\ee
as it should be.

\medskip
\noindent
{\underbar{N = 2}

\smallskip
\noindent
With $\xi: = \xi_1 - \xi_2$ and $x: = x_1 - x_2$ we obtain from (30)
\be
\xi_1 + \xi_2 = {{\dot{x}_1 + \dot{x}_2 - 2v} \over {1 - {g \over 2}
\vert x \vert}}
\ee
and
\be
\xi = {{\dot{x}} \over {1 + {g \over 2} \vert x \vert}}
\ee
According to (35) the constraint (32) leads to an uniform motion of the
center of mass
\be
\ddot{x}_1 + \ddot{x}_2 = 0
\ee
This result confirms the expectation that our gauge fixing (26) together
with the constraint (32) describes particle motion in an inertial frame
for N = 2.\\
Insertion of (36) and (32) into (31) leads to the EOM for $x(t)$
\be
\ddot{x} - {g \over 4}{{\dot{x}^2 \epsilon (x)} \over {~~1 + {g \over 2}
\vert x \vert}} = 0
\ee
This EOM may be derived from the reduced particle Lagrangean
\be
L_{red} = {1 \over 4} {{\dot{x}^2} \over {~~1 + {g \over 2} \vert x \vert}}
\ee
The EOM (37) and $L_{red}$ (39) are invariant with respect to time
translations. Therefore our residual symmetry is the full Galilei group.

It is straightforward to show that $L_{red}$ follows from our total
Lagrangean (10) and (18) by inserting the solutions for the gauge fields.
The EOM (38) may be solved explicitly:\\
With the conserved energy $E$ corresponding to $L_{red}$
\be
E = {1 \over 4} {{\dot{x}^2} \over {~~1 + {g \over 2} \vert x \vert}}
\ee
(38) takes the form
\be
\ddot{x} - g E \epsilon (x) = 0
\ee
This is Newton's EOM for a gravitational potential whose strength is proportional
to $E$.\\
From (40) we conclude, that the relative particle motion described by $x (t)$
is bounded for $E = 0$ and for $E > 0$ with $g < 0$, but unbounded in all
other cases. From (41) we infer that $x (t)$ is given by a second order
polynomial.

\section{Particles on a circle}

Let us apply the foregoing results to a compact manifold, a circle. In
order to do that we have to substitute everywhere
\be
x \to \varphi,~~~~- \pi < \varphi < \pi
\ee
and to require $2 \pi$-periodicity in $\varphi$ for our gauge fields.
In doing so the only problem arises from the step function contained
in the expressions for $F (\varphi, t)$ and $e (\varphi, t)$. It
decomposes into a periodic and a non-periodic part
\be
\epsilon (\varphi) = {2 \over \pi} \sum^\infty_{n=1} {{\sin n \varphi} \over
n} + {\varphi \over \pi}
\ee
Therefore periodicity of $F (\varphi, t)$, which now, as the general solution
of (21), is given by
\be
F(\varphi, t) = - {g \over 2} \sum_\alpha \xi_\alpha (t) \epsilon (\varphi
- \varphi_\alpha (t)) + f_0 (t)
\ee
forces a vanishing coefficient of $\varphi$
\be
\sum_\alpha \xi_\alpha = 0
\ee
in accordance with (23).\\
In order to reproduce the gauge (26) with a periodic gauge function
$\Lambda (\varphi, t)$ we have to put
\be
\Lambda (\varphi, t) = \Lambda_0 (\varphi) - (a + vt)
\ee
with
\be
\Lambda_0 (\varphi): = 2 \arctan (tg {\varphi \over 2})
\ee
leading to
\be
\Lambda_0 (\varphi) = \varphi \qquad {\rm for} - \pi < \varphi < \pi.
\ee
With this choice $e (\varphi, t)$ is given by
\begin{eqnarray}
e (\varphi, t) = & - & v + (\Lambda_0 (\varphi) - (a + vt)) f_0 (t) 
+\nonumber\\
& + & {g \over 2} \sum_\alpha \xi_\alpha (t) \epsilon (\varphi
- \varphi_\alpha (t)) (\Lambda_0 (\varphi) - \varphi_\alpha (t))
\end{eqnarray}
Periodicity of $e (\varphi, t)$ enforces the additional constraint
\be
\sum_\alpha \xi_\alpha \varphi_\alpha = 0
\ee
Thus we obtain instead of (30) the relation
\begin{eqnarray}
\xi_\alpha & = & \dot{\varphi}_\alpha - v + (\varphi_\alpha - (a + vt)) 
f_0\nonumber\\
& + & {g \over 2} \sum_\beta \xi_\beta \vert \varphi_{\alpha \beta} \vert
\end{eqnarray}
with $- \pi < \varphi_\beta < \pi~~~\forall \beta$ and 
$\varphi_{\alpha \beta}: = \varphi_\alpha - \varphi_\beta$.\\
In order to have time-translational invariance for our particle dynamics,
the t-dependence of $\xi_\alpha (t)$ must arise solely from the
particle trajectories $\varphi_\alpha (t)$. Eq. (51) tells us, that this
requires
\be
f_0 = 0
\ee
so that we have for the determination of our trajectories the same set
of equations as in section 4 supplemented by (50).\\
Let us consider an initial condition where
\begin{itemize}
\item[i)] all particles are at different positions at $t = 0$
\be
\varphi_\alpha (0) \not= \varphi_\beta (0) ~~~\forall~~~\alpha \not= \beta
\ee
\item[ii)] the angular velocities at $t = 0$ are equal to $v$ for 
$N-2$ particles, i.e.
\be
\dot{\varphi}_\alpha (0) = v~~~~~~~~~~~~~~~~{\rm for}~~~~~~1 \le \alpha
\le N-2\ .
\ee
\end{itemize}
Then we obtain from (30) - (32) and (50)
\be
\dot{\varphi}_\alpha (t) = v~~~~~~~~~~~~~~~\forall~~\alpha,~~~\forall
t \in {\bf R}^1
\ee
i.e. we have no relative particle motion but only a collective rotation
with constant angular velocity\footnote{It is quite interesting to note 
that for $N = 2$ we can distinguish locally whether the particles are on 
a line or on a large circle.}.\\
The proof of (55) runs as follows:\\
Due to (30) and (50) we may represent $\xi_{1,2}$ as linear combinations
of the others
\be
\xi_{1 \atop 2} = \pm {1 \over {\varphi_{12}}} \sum^N_{\alpha = 3}
\xi_\alpha \varphi_{1 \atop 2} \alpha
\ee
At $t = 0$ we obtain from (54), (56) and (30)$_{1 \le \alpha \le N-2}$
\be
\xi_\alpha = 0~~~\forall \alpha
\ee
which holds, due to (31) $\forall t \in {\bf R}^1$. Inserting (57) for
arbitrary $t$ into (30) leads to the desired result (55).

\section{Conclusions}

We have shown that the application of the gauge principle to point
particles in one dimension leads to a nontrivial interaction between
two particles on a line. On a circle we observe for appropriate
initial conditions a collective rotation only. In this case a sensitivity
analysis for perturbations of the initial conditions would be useful.\\
Work on the gauging of space translations for nonrelativistic matter fields
is in progress. Results will be reported elsewhere.\\
Next we intend to treat the 2d-case with a Chern-Simons interaction for the
gauge fields. Thereby we will include the second central charge of the
Galilei group in\linebreak
\noindent
$(2+1)$-dimensions in the free-particle Lagrangean
(cp.~\cite{Lukierski}).

\bigskip
\leftline{\Large\bf Acknowledgement}
\medskip
I'm grateful to my colleagues Ph.~Blanchard and E.H.~de Groot for fruitful
discussions and suggestions.

\end{document}